\documentclass{ws-p8-50x6-00}

\def\Journal#1#2#3#4{{#1} {\bf #2}, #3 (#4)}

\def\PLB{{\em Phys. Lett.}  B}

\def\EPJ{{\em Eur.Phys.J.} C}

\begin{document}

\title{New Results on $\alpha_s$ and Optimized Scales}

\author{S. Bravo}

\address{Institut de F\'{\i}sica d'Altes Energies, Universitat Aut\`onoma de Barcelona, 
\\E-08193 Bellaterra, Spain,
\\E-mail: bravo@ifae.es}


\maketitle

\abstracts{
 A summary of the latest $\alpha_s$ results at LEP1 and 
LEP2 from event-shape predictions at ${\cal O}(\alpha_s^2)$+NLLA is presented. 
Later these are compared to measurements obtained using the Experimentally 
Optimized Scale method. Finally the $\alpha_s$ measurement from  the 4-jet 
rate is discussed.}

\section{Introduction}

The understanding of the role played by the renormalization scale parameter $\mu$ in the $\alpha_s$ 
measurements is the main goal of the coming sections. Such a parameter appears in the 
perturbative series of the QCD predictions, which for any observable is independent of 
this un physical parameter 
if all the orders are known. However, usually only the LO and NLO terms have been 
calculated, and for some observables also the resummation of large logarithms (NLLA) exists.
 The truncated perturbative prediction is then a function of the renormalization scale.

\section{Latest $\alpha_s$ results at LEP1 and LEP2 from event shapes}

In this section results of $\alpha_s$ measurements 
by the four LEP collaborations from 2- and 3-jet observables are summarised. 
For such observables the theoretical predictions are known at 
least at ${\cal O}(\alpha_s^2)$, i.e.,

\begin{equation}
\begin{array} {rcl}
\frac {1}{\sigma_{tot}} \frac {d\sigma \left ( O_3 \right )} {dO_3} &=&
\frac{\alpha_s \left( \mu \right )}{2\pi} A \left ( O_3 \right ) +
\left ( \frac {\alpha_s \left( \mu \right )}{2\pi} \right )^2 
\left [ A \left ( O_3 \right ) 2 \pi b_0 \ln \left (\frac{\mu^2} {s} \right )
+ B \left ( O_3 \right ) \right ] \\
&&{} + f \left ( \alpha_s^n \ln^m O_3 \right ) 
\end{array}
\label{eq:O3}
\end{equation}

\noindent where {\it s} is the centre-of-mass energy, $O_3$ a 3-jet observable 
and f groups the NLLA terms.

\subsection{LEP1 results}

LEP1 measurements were produced using data taken in 1991-1995, 
when LEP was running at the Z peak. In figure \ref{fig:lepresults}(a) results 
from the four LEP collaborations\cite{Lep12A}$^-$\cite{Lep12O} using
different sets of event shapes are summarised. The renormalization scale was 
fixed to the centre-of-mass energy ($M_Z$), and its uncertainty was 
estimated by varying $x_{\mu}=\frac {\mu}{M_Z}$ between 0.5 and 2. 
This uncertainty dominates the total error, being about 80\% of it. 
The measurements are fully compatible within errors.

\subsection{LEP2 results}

\begin{figure}[hb]
\centering
\mbox{\subfigure[LEP1]{\epsfig{figure=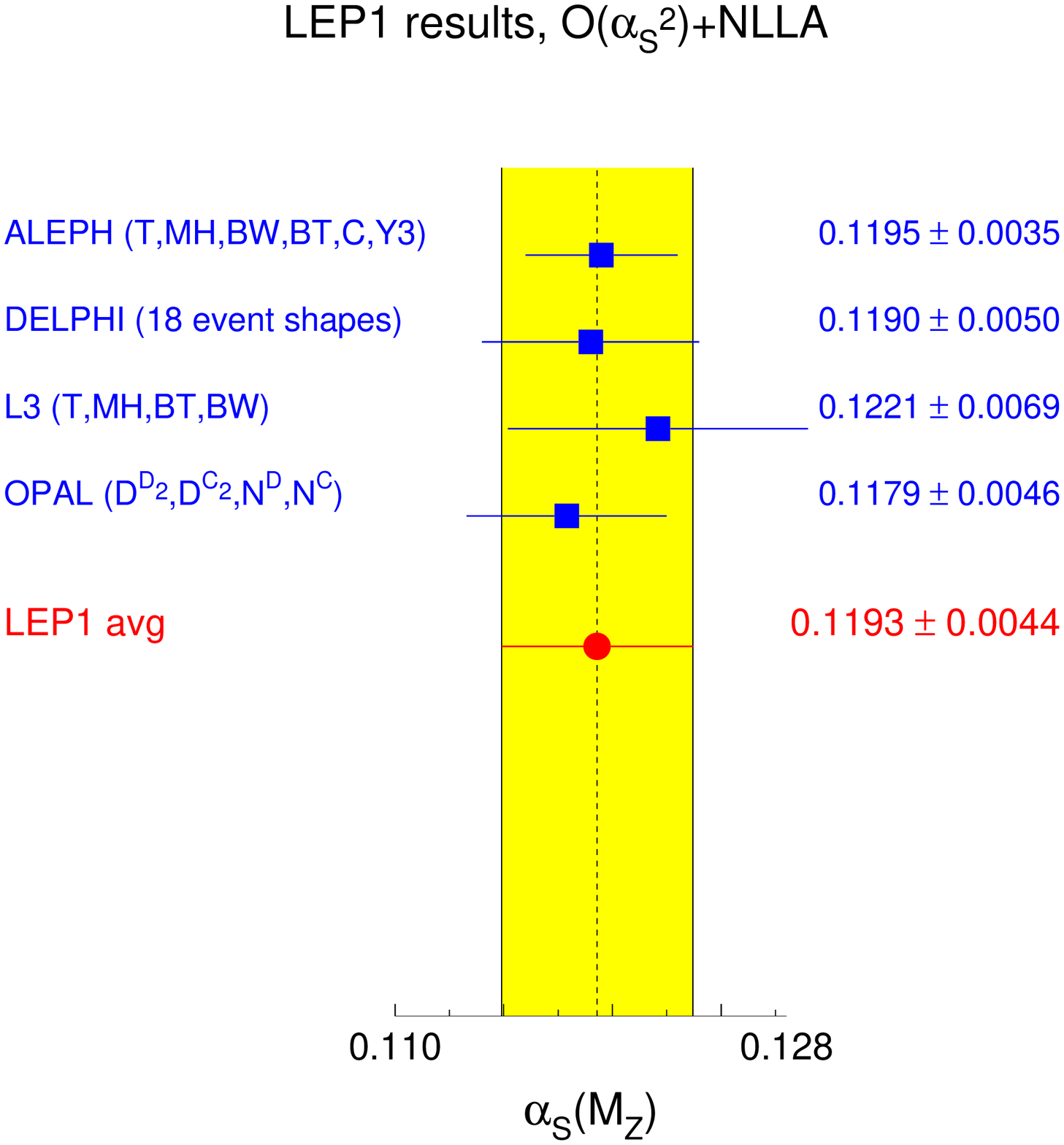,width=.41\textwidth}} \quad
\subfigure[LEP2]{\epsfig{figure=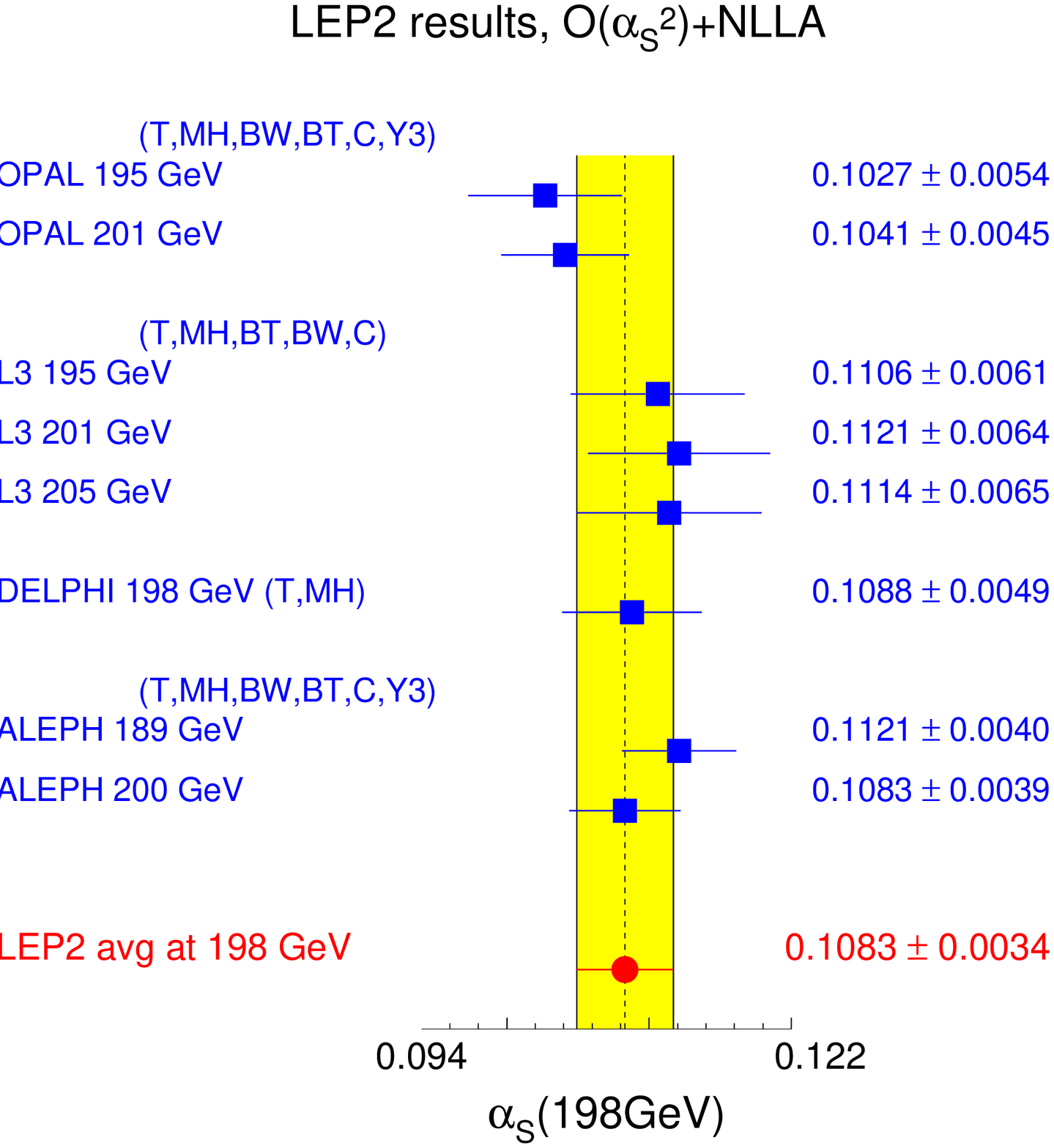,width=.41\textwidth}} }	
\caption{LEP1 and LEP2 results for $\alpha_s$ measurements from the four LEP collaborations. The average was 
obtained using the rescaling method$^5$} \label{fig:lepresults} 
\end{figure}

At LEP2 with data at higher energies, smaller 
hadronization uncertainties are expected, and a different background treatment is 
applied in the analyses. Figure \ref{fig:lepresults}(b) shows some results from the four LEP 
collaborations\cite{Lep12A}$^,$\cite{Lep2D}$^,$\cite{Lep2L}$^,$\cite{Lep12O}. The scale uncertainty 
still dominates the total error, leading to $\alpha_s(M_Z)$ measurements in good 
agreement with the ones obtained at LEP1.

\section{Optimized Scales}

For most event shapes the calculations of the perturbative series exist only for the first
two terms. The ratio of the NLO contribution with 
respect to the LO one is used to estimate the importance of the unknown terms. In many cases 
this ratio is close to unity, an indication of the poor convergence of the perturbative series. 

One can think of a value of the renormalization scale chosen in order to match the theoretical
predictions to data. Such an optimal scale is found, without any theoretical assumption, 
by a combined fit of $\alpha_s$ and $x_{\mu}$. 
This is the so called Experimentally Optimized Scales method (EOS), from which results 
follow in the coming sections. The optimized scale can differ for different observables, as 
the convergence of the truncated series does not have to be the same.

\subsection{Optimized Scales results in DELPHI}

\begin{figure}[th]
\begin{center}
\begin{tabular} {cc}
\epsfxsize4.8cm
\epsfysize5.8cm
\figurebox{4.8cm}{5.8cm}{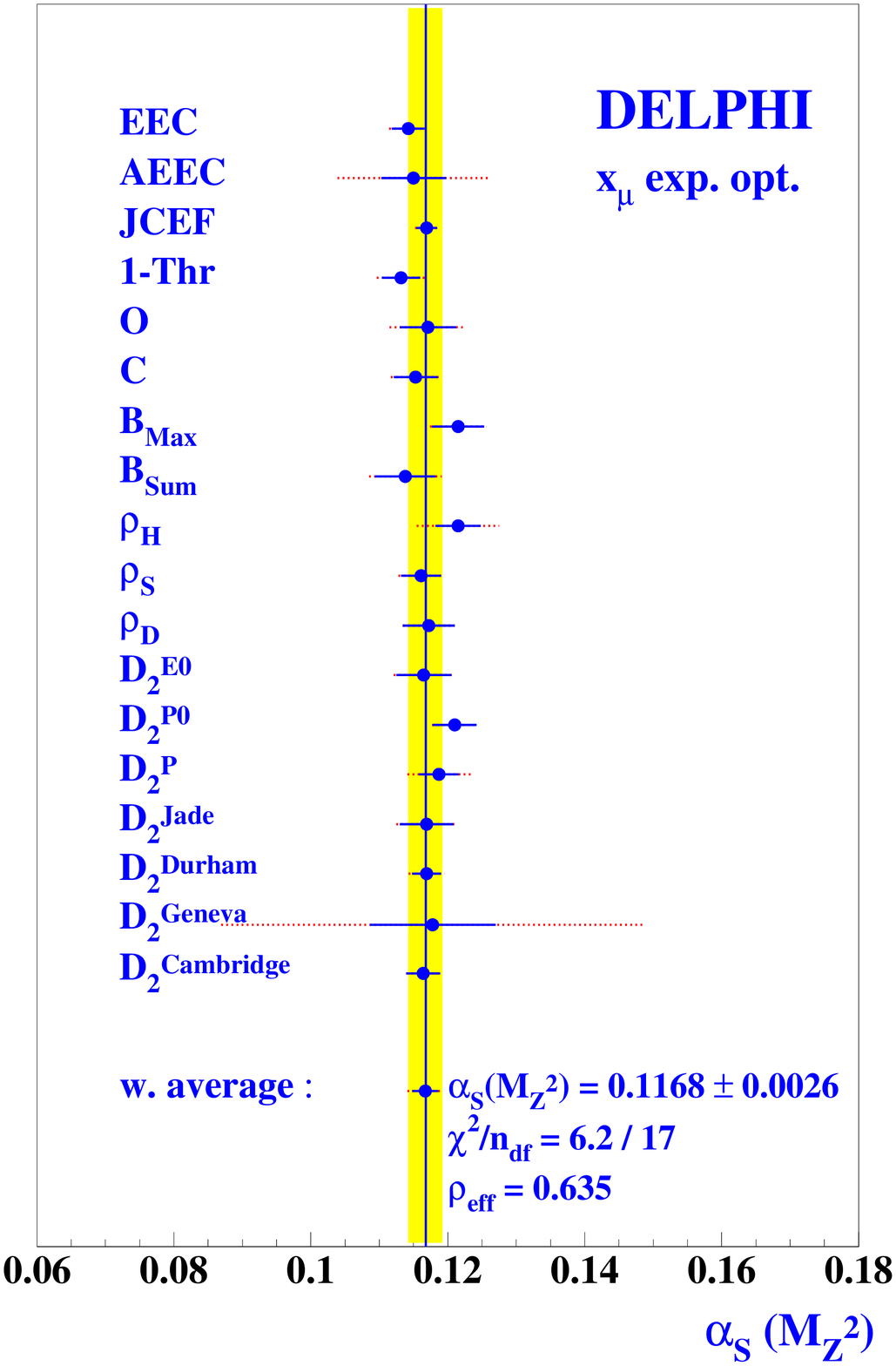} &
\epsfxsize4.8cm
\epsfysize5.8cm
\figurebox{4.8cm}{5.8cm}{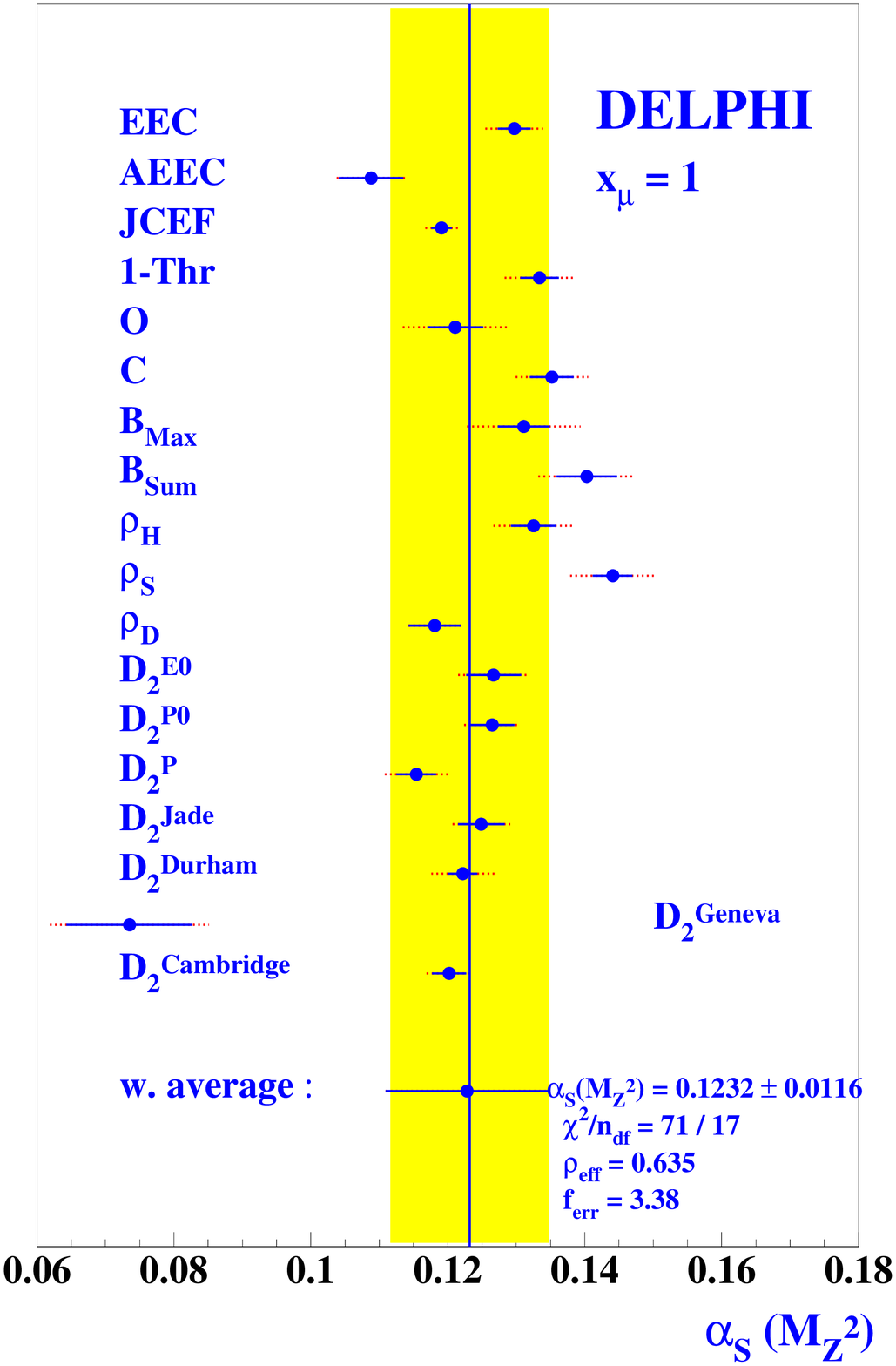} 
\end{tabular}
\caption{DELPHI results with the EOS and ${\cal O} (\alpha_s^2)$ methods.}
\label{fig:exp_opt}
\end{center}
\end{figure}

\begin{figure}[th]
\epsfxsize5cm
\epsfysize6cm
\figurebox{5cm}{6cm}{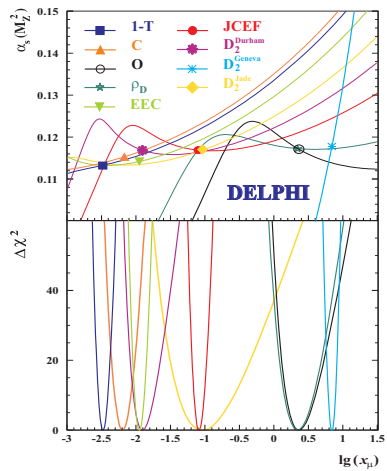}
\caption{Scale dependence for different observables in the EOS method.} 
\label{fig:xmurange}
\end{figure}

DELPHI has recently updated a LEP1 study on the EOS at ${\cal O} (\alpha_s^2)$ method using a set of 16 event-shape observables\cite{Lep1D}. Results are compared to fits using 
 ${\cal O} (\alpha_s^2)$ and ${\cal O} (\alpha_s^2)$+NLLA predictions. In figure~\ref{fig:exp_opt} the 
dispersion of the fitted $\alpha_s$ is shown to be much smaller for EOS at 
${\cal O} (\alpha_s^2)$. Furthermore, in EOS the uncertainty due to 
hadronization corrections becomes the largest, since the scale uncertainty is heavily reduced. 
The scale uncertainty is measured in EOS as the largest deviation in $\alpha_s$ when $x_{\mu}$ 
is varied between 0.5 and 2 times the experimentally optimized value.  EOS at 
${\cal O} (\alpha_s^2)$, following the 
DELPHI conclusions, has then a small scale uncertainty with the 
total error heavily reduced. In figure \ref{fig:xmurange} the large dispersion of the 
experimentally optimized scales is shown, going from $x_{\mu}$ (here defined as $\frac{\mu^2}{M_Z^2}$)
around 0.003 to 7.10, i.e. $\mu$ from 5 GeV to 240 GeV.

Other conclusions drawn by DELPHI are that EOS at ${\cal O} (\alpha_s^2)$ 
describes the data over the whole fit range better than resummed predictions. 
Average results from the 16 observables show a good 
agreement between the EOS method and the fits to resummed 
predictions as seen in table~\ref{tab:asresults}.

\begin{table}[t]
\begin{center}
\begin{tabular}{|c|c|c|}
\hline
{} & $\alpha_s$ & Total Error \\
\hline
EOS & $0.117$ & $0.003$ \\
${\cal O}(\alpha_s^2)$ + NLLA & $0.119$ & $0.005$ \\
\hline
\end{tabular}
\end{center}
\caption{$\alpha_s$ results from DELPHI. \label{tab:asresults}}
\end{table}

The study also includes results obtained when choosing the 
optimal scale according to some theoretical assumption (such as vanishing 
NNLO terms). A larger dispersion in the fitted $\alpha_s$ is found, but results
are fully compatible with EOS. The study 
concludes that the best method for $\alpha_s$ extraction from 2- and 3-jet 
observables is EOS at~${\cal O} (\alpha_s^2)$.

\subsection{Other Optimized Scales results}

A recent analysis by OPAL\cite{Lep12O} has 
lead to different conclusions (and results from SLD\cite{SLD} confirm such 
discrepancies). They show that ${\cal O}(\alpha_s^2)$ predictions describe better the 
data if the scale is also fitted. However, one can not arrive to a conclusion if a 
comparison with respect to resummed predictions is done, as the best prediction depends
 on the observable. Following this analysis resummed predictions have a smaller $x_\mu$ 
dependence, and therefore the smallest scale uncertainty. They show that the shape of both 
$\alpha_s$ and $\chi^2$ depend strongly on the scale, but with a stable minimum. Following 
OPAL's studies the best method for $\alpha_s$ extraction 
from 2- and 3-jet observables would be EOS at ${\cal O} (\alpha_s^2)$ +NLLA.

\section{$\alpha_s$ from the 4-jet rate}

In this section the first $\alpha_s$ measurement from the 4-jet rate is summarised\cite{4jetrate}. 
These observables have an attractive characteristic. Since the LO term for 4-jet observables is proportional 
to $\alpha_s^2$, these observables have less sensitivity to possible large sources of systematics, as 
$\frac{\Delta\alpha_s}{\alpha_s}=\frac{1}{2}\frac{\Delta\sigma}{\sigma}$.

\begin{figure}[h]
\epsfxsize6.0cm
\epsfysize5.5cm
\figurebox{6.0cm}{5.5cm}{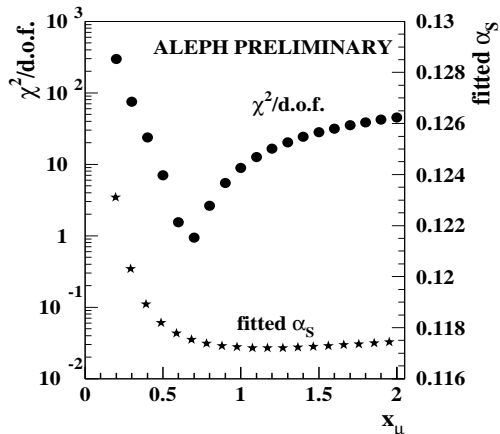}
\caption{$\alpha_s$ and $\chi^2$ dependence on the renormalization scale for 
the 4-jet rate fit . } 
\label{fig:xmu4jet}
\end{figure}

The measurement is done by fitting ${\cal O} (\alpha_s^3)$ +NLLA prediction for the 4-jet rate to
 1994 ALEPH data corrected to parton level. 
The results when fitting only $\alpha_s$ and when doing a combined fit of $\alpha_s$ and 
$x_{\mu}$ (i.e. EOS at ${\cal O} (\alpha_s^3)$ +NLLA) can be seen in table~\ref{tab:asresults4jet}, where errors are statistical only. In figure~\ref{fig:xmu4jet} a 
strong dependence of the $\chi^2$ with the renormalization scale 
is observed, but with a clear minimum around 0.7. However, the $\mu$ dependence of the fitted 
 $\alpha_s$ is 
much smaller, showing that the 4-jet rate has a small scale uncertainty. The preliminary result was : 
\begin{center}
$\alpha_s(M_Z)=0.1172 \pm 0.0001_{stat} \pm 0.0003_{exp} 
\pm 0.0008_{had} \pm 0.0040_{theo}$. 
\end{center}

\begin{table}[bh]
\begin{center}
\begin{tabular}{|c|c|c|}
\hline
{} & $\alpha_s$ & $x_{\mu}$\\
\hline
${\cal O}(\alpha_s^3)$ + NLLA & $0.11724 \pm 0.00012$ & $1.$ \\
EOS ${\cal O}(\alpha_s^3)$ + NLLA & $0.11760 \pm 0.00014$ & $0.67 \pm 0.03$ \\
\hline
\end{tabular}
\end{center}
\caption{Results from the 4-jet rate by ALEPH. \label{tab:asresults4jet}}
\end{table}

\noindent The theoretical error should be reduced in the final results. It was estimated by the 
linear sum of the scale and NLL uncertainties (0.0016 and 0.0024, respectively). So the 4-jet rate has been 
proven to be a good observable for $\alpha_s$ measurements, since
still preliminary results have already reached the precision of previous measurements which used  2- and 3-jet observables.

\section{Conclusions} 
The latest $\alpha_s$ results at LEP1 and LEP2 have been summarised, showing that 
measurements at different energies lead to compatible $\alpha_s(M_Z)$ results. Then two 
analyses by DELPHI and OPAL on Experimentally Optimized Scales have been discussed, 
showing some discrepancies in the final conclusions. However, they both 
agree on the reduction of the renormalization scale uncertainty when a combined fit of both
$\alpha_s$ and $\mu$ is done. Finally, the first 
$\alpha_s$ measurement from the 4-jet rate has been briefly presented, with preliminary
results in agreement with previous 2- and 3-jet measurements, and with similar errors.

\section*{Acknowledgments}
I would like to thank Guenther Dissertori for both the discussions and support
 during the preparation of the conference talk.


\begin{thebibliography}{99}
\bibitem{Lep12A} ALEPH Collaboration. ALEPH 2000-044. CONF 2000-027

\bibitem{Lep1D}  DELPHI Collaboration. \Journal{\EPJ}{14}{557}{2000} 

\bibitem{Lep1L}	 L3 Collaboration. \Journal{\PLB}{489}{65}{2000}

\bibitem{Lep12O}  The JADE and the OPAL Collaboration. \Journal{\EPJ}{17}{19}{2000} 

\bibitem{schmelling} M. Schmelling. hep-ex/0006004, 2000

\bibitem{Lep2D}	   J. Drees {\it et al}. DELPHI 2000-116. CONF 415

\bibitem{Lep2L}    L3 Collaboration. L3 Note 2555. Submitted to ICHEP2000

\bibitem{SLD} P.N. Burrows {\it et al.} \Journal{\PLB}{382}{157}{1996}

\bibitem{4jetrate} ALEPH Collaboration. ALEPH 2000-045. CONF 2000-028 


\end{thebibliography}
\end{document}